----------
X-Sun-Data-Type: default
X-Sun-Data-Description: default
X-Sun-Data-Name: FHsissa.tex
X-Sun-Content-Lines: 597

\documentstyle[12pt]{article}
\def\eq{\begin{equation}}
\def\en{\end{equation}}
\newcommand \be  {\begin{equation}}
\newcommand \bea {\begin{eqnarray} \nonumber }
\newcommand \ee  {\end{equation}}
\newcommand \eea {\end{eqnarray}}

\def \bi{\bibitem}

\def\d{{\rm d}}
 \def\(({\left(}
 \def\)){\right)}
\def\bi{\bibitem}

\def \ov{\over}

\def \d{{\rm d}}
\def \e{{\rm e}}

\def \nn{\nonumber}
\def \beqna{\begin{eqnarray}}
\def \eeqna{\end{eqnarray}}
\def \beq{\begin{equation}}
\def \eeq{\end{equation}}
\def \be{\begin{equation}}
\def \ee{\end{equation}}
\def \ln{{\rm ln}}
\def \ov{\over}

\def \ol{\overline}

\def \r{\right}

\def \l{\left}
\def \ln{{\rm ln}}
\def \la{\langle}
\def \ra{\rangle}
\def \Tr{{\rm Tr}}

\def \ab2{\alpha\beta^2}
\def \s{\sigma}
\def \la{\langle}
\def \ra{\rangle}
\def \hpsi{\hat{\psi}}
\def \cL{{\cal L}}
\def \cD{{\cal D}}
\def \tq{\tilde{q}}

\begin{document}

\title{Glassy transition and aging in a model without disorder.}
\author{ Silvio Franz and John Hertz\\
\\
  NORDITA, Blegdamsvej 17, DK-2100 Copenhagen \O, Denmark\\
}
\date{August 1994}
\maketitle

\begin{abstract}
We study the off-equilibrium relaxational dynamics of the Amit-Roginsky
$\phi^3$ field theory, for which the mode coupling approximation is exact.
We show that complex phenomena such as aging and ergodicity breaking
are present at low temperature, similarly to what is found in long range
spin glasses. This is a
%first
%step towards the
 generalization of
mode coupling theory of the structural glass transition to off-equilibrium
situations.
\end{abstract}
\vspace{.5 cm}
PACS 6470P, 7510N
\vspace{.5cm}
Preprint NORDITA 94/39, cond-mat/9408079

Recently there has been important progress in the understanding of
the off-equilibrium dynamics of disordered systems.
Although aging phenomena were discovered experimentally in spin glasses
in 1983 \cite{AGESWE} and subsequently studied by several groups in spin
glasses \cite{AGESG} and many other different disordered systems, only
quite recently has a dynamical mean field theory (DMFT)
for non-equilibrium phenomena, based on microscopic spin glass models,
been put forward \cite{cuku1,frme,cukusk}. Among the most interesting
feaatures which emerge from the theory are the violations at low temperature
of time translation invariance (TTI) and of the fluctuation-dissipation
theorem (FDT), even for very large times, implying breaking of ergodicity.
This breaking of ergodicity can not be described simply as a separation
of the phase space into different ergodic components within which the
system thermalizes as in usual broken-symmetry phases.  Rather, the
distribution function continues to spread over larger regions of
configuration space as time goes on.  This spreading gets slower and
slower but never stops; thus the system never comes to equilibrium,
and its properties depend on the time elapsed since the
quench to low temperature.

Far from being confined to the realm of disordered system aging phenomena
have been observed in a number of different systems without disorder, in
experiments \cite{struik}
and numerical simulations \cite{binder}.

At present
there
is a great deal of theoretical research  on
spin-glass like phenomena in non-random systems
\cite{boumez,mapa1,kuri}. In \cite{boumez,mapa1}
 a mechanism has been proposed in which
an effective disorder can be induced in {\it a priori} uniform
 systems (althoug with "complicated" interactions among the variables).
In this paper we focus on the dynamical mean field theory for glasses,
usually called mode coupling theory (MCT), showing how it is possible to extend
it in order to include  non equilibrium phenomena.

It has been noted \cite{hor} that
the mean field dynamical equations of many disordered mean field models
have, in the high temperature phase, a structure which is surprisingly
similar to that of the equation of the mode coupling theory
  often used to
describe the structural glass transition \cite{leut,gotze}.
The mode coupling equations have been derived from molecular
or hydrodynamical models based on an approximation
in  which vertex correction are neglected in the perturbation expansion of
the self-energy \cite{dasmaz}.
 Systematic use is made of the hypothesis that the system
under study is at thermal equilibrium.
On the other hand there is experimental evidence that real glasses
are out of equilibrium \cite{gotze}. Even in MCT the ideal
glass transition is described as
one into a non-ergodic state.   Thus current mode coupling theory can at
best give a good description of the physics at high enough temperature.

This letter is a step
towards the extension of the MCT below the glassy temperature, including
the possibility of studying nonequilibrium phenomena.
Technically, we construct a theory in which we do not assume TTI  or the
FDT {\it ab initio} in the derivation of the equations.
As an example we have chosen to study a simple  model without quenched
disorder, where
the mode coupling approximation is exact.
This is the Amit-Roginsky $\phi^3$ field theory \cite{amro}, originally
proposed to study the critical properties of the 3-state Potts model.
We will find dynamical equations identical to that of the spherical
{\it p}-spin glass
model \cite{criso,hor,cuku1}
 for $p=3$. At high temperature, supposing TTI and FDT one gets
usual MCT equations, similar to those first proposed by Leutheusser to
study the glassy transition. At low temperature there is a phase where
ergodicity is broken in a way now known from mean field spin glass
dynamics.

Amit and Roginsky (AR) consider an $O(3)$ symmetric $\phi^3$ field theory
where the basic fields transform under rotation according to representations
of high angular momentum quantum number $l$.
Recently the same method has been exploited in two
interesting papers \cite{mouwe}, to find soluble limits of nonlinear
stocastic equations. The method is applicable any time the nonlinearity is
represented by a bilinear term.  In this letter we discuss for simplicity
the dynamics of the original AR model. In a forthcoming paper \cite{fh2}
we will discuss the case of the fluctuating compressible hydrodynamics of
Das and Mazenko \cite{dasmaz}, which, although formally much more complicated
than the present model, gives rise to equations of with the same structure.

The AR model is defined by the $O(3)$ invariant Lagrangian
\beqna
{\cal L}[\psi,\psi^*]&=& \sum_{m=-l}^l \psi_m^*(x)[\mu^2 -\nabla^2]\psi_m(x)
\nonumber
\\
&+&{g\ov 3!}\sqrt{N}\sum_{m_1,m_2,m_3}
\l(
\begin{array}{lll}
l & l & l \\
m_1 & m_2 & m_3
\end{array}
\r)
[\psi_{m_1 }(x)\psi_{m_2 }(x)\psi_{m_3 }(x) +c.c.]
\label{L}
\eeqna
with $N=2l+1$, and $\psi_m$ transforms according to the irreducible
$N-$dimensional representation of $O(3)$. In the following we will
denote by $\cL_I[\psi,\psi^*]$ the cubic part of the lagrangian.
The coefficients
$\l(
\begin{array}{lll}
l & l & l \\
m_1 & m_2 & m_3
\end{array}
\r) $
are the Wigner 3$j$-symbols. To have a non vanishing interaction term
$l$ must be even. Amit and Roginsky have shown that in the limit
$l\to \infty$ vertex corrections in the perturbation expantion
can be neglected.
Thus, defining the equilibrium corrlation function $G(x,y)=\sum_m\la
\psi_m(x)\psi^*_m(y) \ra_{Gibbs} $, the perturbative
series can be resummed, and
the Dyson equation of the model in real space is simply:
\be
G^{-1}(x,y)=G_0^{-1}(x,y)-{g^2\ov 2}[G(x,y)]^2,
\label{Tyson}
\ee
where $G_0^{-1}(x,y)=\delta(x-y)(\mu^2-\nabla^2)$.
Similarly one finds that
the free energy as a functional of $G$ can be written
\be
F=-\Tr {\rm Log}(G)+\int \d x \d y G_0^{-1}(x,y)G(y,x)
-{g^2\ov 6}\int \d x \d y G(x,y)^3
\label{free}
\ee
and (\ref{Tyson}) is its associated variational equation.
It is assumed in deriving (\ref{free}) that $\la \psi_m(x)\ra = 0$.

In order to
avoid problems connected with the unnormalizability of the Gibbs measure
for a $\phi^3$ theory, we will consider a ``spherical" variation
of the AR model, imposing the constraint
 $\sum_m \psi_m(x)\psi_m^*(x)=N\tq$.
In this situation, $\mu^2$ can be viewed as a Lagrange multiplier
which enforces the constraint. Eq. (\ref{Tyson}) for $x=y$ should
in this case be read as an equation for $\mu^2$.

We study the relaxational  dynamics of this model, restricting
our attention
to the zero-dimensional case (we drop the $x$ dependence and the gradient
term in (\ref{L})). The
extension of our results to finite dimension is immediate.
We consider the Langevin equation
\be
{\d\ov \d t}\psi_m(t)=-\mu^2(t)\psi_m(t)-{\partial {\cal L}_I\ov \partial
\psi_m^*} +\eta_m(t)
\label{lang}
\ee
where $\eta_m$ is a white noise with correlation function
$\la \eta_m(t)\eta_n^*(s)\ra =2T\delta_{mn} \delta(t-s)$, and $T$ is the
temperature.
We will see {\it a posteriori} that it is possible to choose $\mu$ as a
function of $t$ to impose the spherical constraint at all times.
In order to generate the perturbation expansion of eq.(\ref{lang}),
we consider the Martin-Siggia-Rose generating functional
\be
Z[h]=\int {\cal D}\psi{\cal D}\psi^*
\l\la
   \prod_{m,t}\delta
\l(
{\d\ov \d t}\psi_m(t)+{\partial {\cal L}_t\ov \partial \psi_m^*} -\eta_m(t)
\r)
\r\ra
\exp\l(\int \d t h_m^*(t)\psi_m(t)\r)
{\cal J}[\psi]
\label{Z}
\ee
${\cal J}[\psi]$ is the functional determinant of $\d/\d t+
\partial^2 {\cal L}_t/\partial \psi_m \partial \psi_m^*$ which, with the Ito
convention, is equal to 1, and we denoted by ${\cal L}_t$ the time dependent
``Lagrangian" ${\cal L}_t=\mu^2(t)\sum_m \psi^*_m \psi_m +{\cal L}_I$.
 We now follow a standard procedure \cite{ZJ} to prove
that the structure of the diagrammatic expansion for dynamics is the
same as that for statics.
We introduce an auxiliary field $i\hpsi_m(t)$, to get
$Z=\int {\cal D}\psi{\cal D}\psi^* {\cal D}\hpsi{\cal D}\hpsi^*
\exp
\l(
-\int \d t i\hpsi_m^*\right.$
$\left. [\dot{\psi_m}-i\hpsi_m +(\partial \cL /\partial \psi_m^*)]
\r).$
For simplicity we have set $h_m=0$. This will not affect our subsequent
manipulations.
We then introduce an even grassman variable $\theta$, and the
``superfield"
$\phi_m(t)=\psi_m(t)+\theta i\hpsi_m(t)$. In terms of
these, $Z$ reads,
\be
Z=\int \cD \phi \cD \phi^*
\exp
\l(
- \int \d t \int \d \theta \{ \phi_m^*(t,\theta)[(1-
\theta{\partial\ov \partial
\theta}){\partial\ov\partial t}-T\theta{\partial\ov \partial
\theta}+\mu^2(t) ]\phi_m(t,\theta)
+\cL_I[\phi,\phi^*]\}
\r)
\label{zzz}
\ee
The time integration here should be intended to start from an initial
time $t=0$, where the system is in a random initial condition. This
corresponds to the experimental situation of a sudden quench of
the system  from very high
temperature at the initial time.

The interaction term in (\ref{zzz}) is represented
by the same function $\cL_I$ as in
statics, but its arguments are now the superfields $\phi$ and $\phi^*$
rather than just of $\psi$ and $\psi^*$.  Thus the structure of
the dynamical perturbation expansion in terms of
the superfield is formally  identical to that of statics in
terms of the field.  We define at this point the correlator
\beqna
G(t,\theta;t',\theta')&=&\la \phi_m(t,\theta)\phi_m^*(t',\theta') \ra
\nn
\\
&=& \la \psi_m(t)\psi_m^*(t') \ra +\theta \la i\hpsi_m(t)\psi_m^*(t') \ra
\nn
\\
&+& \theta' \la \psi_m(t)i\hpsi_m^*(t') \ra +\theta\theta'
\la i\hpsi_m(t)i\hpsi_m^*(t') \ra.
\eeqna
We see that $G$ codes for the correlation function
$C(t,t')=\la \psi_m(t)\psi_m^*(t') \ra$ and the response function
$r(t,t')=\la \psi_m(t)i\hpsi_m^*(t') \ra =\la \delta \psi_m(t)/\delta
\eta_m(t')\ra$, while the conservation of probability implies
$\la i\hpsi_m(t)i\hpsi_m^*(t') \ra=0$. Causality and the Ito convention imply
$r(t,t')=0$ for $t'>t$ and $r(t^+,t)=1$.
The zeroth order correlator in dynamics is defined by $G_0^{-1}=(1-
\theta{\partial\ov \partial
\theta}){\partial\ov\partial t}-T\theta{\partial\ov \partial
\theta}+\mu^2(t)$.
In terms of $G_0$ and $G$ defined in this way, the Dyson
equation of the theory has the same form as
(\ref{Tyson}), replacinging
$x$ and $y$ respectively by $(t,\theta)$ and $(t',\theta')$. The appearence
of $C$ and $r$ only in the combination given by $G$ is a consequence of the
gradient character of the Langevin equation.

Inverting the equation (\ref{Tyson}), multipling it
by $G$, and disentangling the
superfield notation, we get coupled equations for the correlation and the
response function; for $t>t'$ we have:
\beqna
{\partial C(t,t')\over \partial t}&=&-\mu^2(t)C(t,t')
+{1\ov 2}g^2
\l\{
    \int_0^t \d s \, 2C(t,s)r(t,s)C(t',s) +\int_0^{t'} \d s C^2(t,s)r(t',s)
\r\}
\nn
\\
{\partial r(t,t')\over \partial t}&=& -\mu^2(t) r(t,t')
+{1\ov 2}g^2
    \int_0^t \d s \, 2C(t,s)r(t,s)r(s,t')
\label{eq1}
\eeqna
 while for arbitrary  $\mu^2(t)$, $C(t,t)$ satisfies:
\be
{1\ov 2}{\d C(t,t)\ov \d t}=-
\mu^2(t)C(t,t)
+{1\ov 2}g^2
    \int_0^t \d s \, 3C^2(t,s)r(t,s) +T.
\label{eq2}
\ee
As mentioned above, we can choose $\mu(t)$ to satisfy
$\sum_m \psi_m^*\psi_m =N \tq$ at all times.
We recognize in (\ref{eq1},\ref{eq2})  well studied equations in mean field
spin glass theory: they are identical to those found in the dynamics of the
so called {\it spherical p-spin model} \cite{hor}
for $p=3$. This is defined by the hamiltonian
\be
H=-\sum_{i<j<k}
J_{ijk} S_i S_j S_k
\ee
where the ``spins" $S$'s are real veriables satisfing the constraint
$\sum_{i=1}^N S_i^2=N\tq$, and the couplings $J_{ijk}$
are independent Gaussian variables,
symmetric under interchange of any pair of indices,
with zero mean and variance $\ol{J_{ijk}^2}=
3g^2/N^2$. The equivalence for large $N$
of the $O(3)$ symmetric vertex with a random
one in dynamics had already been noted in \cite{mouwe}.
Note that eq. (\ref{eq1},\ref{eq2}) are causal first order
integro-differential equations which admit  a
unique solution  for finite times.

In a seminal paper \cite{cuku1} Cugliandolo and Kurchan have found an
asymptotic solution to the equation of the $p$-spin model, showing the
existence of a low temperature phase in which TTI and FDT are violated
even in the infinite time limit. The explicit numerical integration of
equations with the same structure obtained for a related model
\cite{frme} support the asymptotic analysis of \cite{cuku1}, which we
review briefly here in the  present context.
At high temperature, the asymptotic dynamics is completely characterized
by the two functions $C_{as}(\tau)=lim_{t\to\infty}C(t+\tau,t)$ and
$r_{as}(\tau)=lim_{t\to\infty}r(t+\tau,t)$ related by FDT, and such that
$\lim_{\tau\to\infty}C_{as}(\tau)=0$.
The long time limit of $\mu$ is given by  $\mu^2 =(g^2/2T)\tq^2+T/\tq$.
This is just the zero-dimensional limit of eq. (\ref{Tyson}) for the statics.
Accordingly the  ``energy" of the system, $E=\la \cL_I(t)\ra
=-(g^2/3T)\int_0^t \d s C^2(t,s)r(t,s)$ tends to the static value
$E_{static}=-g^2\tq^3/(3T)$. The set of equations (\ref{eq1},\ref{eq2}) reduces
to
a single one e.g. for the response function, almost
identical to the one originally proposed by Lautheusser  \cite{leut}.

Below the critical temperature $T_c= g (\tq/2)^{(3/2)}$,
such a  regime still exists for $t, t' \to \infty$ with $\tau/t$ and $\tau/t'
\to 0$, but with
$\lim_{\tau\to\infty}C_{as}(\tau)=q\ne 0$.  In addition, there is an aging
regime, in which TTI and FDT are violated. In this regime
the limit $t,t'\to \infty$ is taken fixing the ratio
$h(t')/h(t)=\e^{-\s}$ ($\sigma \ge 0$).\footnote{$h(t)$ is an
arbitrary increasing function which
the present theory is not able to predict. On a numerical basis it was claimed
in \cite{cuku1} that $h(t)=t$, leading to the suggestive result
$C(t,t)=C(t'/t)$.}
Here one has
\beqna
C(t,t')&=& \hat{C}(\s)
\nn
\\
Tr(t,t')&=&  u\frac{\partial C(t,t')}{\partial t'}
= -u\l({\d \ln(h(t'))\ov \d t'}\r) {\partial \hat{C}(\s)\ov \partial \s}
\eeqna
Continuity between the homogeneous and aging regimes, implies
$\hat{C}(0)=q$. The value of $q$  is specified at low temperature
by the largest root of the equation
\be
q={T^2\ov g^2}{1\ov (\tq-q)^2}
\nn
\ee
while $\mu^2$ and $u$ are given by
\be
\mu^2={1\ov 2T}g^2(\tq^2-q^2)+{T\ov (\tq-q)};
\;\;\;\;\;\; u={\tq-q\ov q}
\label{spe}
\ee
It is worth noticing that the former equations can be found in the replica
formalism for the {\it p}-spin glass by imposing the condition of marginal
stability.
The energy takes contribution both from the asymptotic and the
aging regins of times, and tends to the limit $E_{Dyn}=-g^2(\tq^3 -(1-u)
q^3)/3T$,
different from its static counterpart.
Figure 1 shows the results of numerical integration of eqs.\ (\ref{eq1})
and (\ref{eq2}) at a temperature where aging occurs.

We have seen that the dynamics of the spherical AR model quite unexpectly
displays glassy behaviour, with a sharp phase transition from
ergodic to non-ergodic behaviour.
For a correct picture of the model at low temperature, a set of coupled
equations for the
correlation and response functions have to be solved.
It has been often discussed in the literature whether glasses undergo a real
phase transition, or whether a better picture is one of a progressive
freezing \cite{dasmaz,gotze}.
MC equations  which fit better with this scenario have been proposed
\cite{dasmaz,gotze}. Even though in this case one can not expect aging
behaviour
 for infinite times
no matter how the limit is taken, the relaxation to equilibrium can be
extremely slow. On an  experimental time scale aging phenomena
(or eventually "interrupted aging") can be present, and the use of the
off-equilibrium
equations can be more appropriate than that of the asymptotic one.

A natural generalization of the MC eq. (\ref{Tyson}), and consequently of
(\ref{eq1},\ref{eq2}) is obtained by replacing
the non linear term by some generic "self-energy" function
$\Sigma (G(x,y))$. This can be introduced on a purely phenomenological
ground, as it has been done by G\"otze in the equilibrium case \cite{gotzetto},
  or derived from some microscopic theory.  Examples of equations of this form
have been  analyzed extensively in
\cite{frme,cukusk}
 in the context of mean field disordered systems. Depending on the form of
$\Sigma$, the long time aging regime
can assume different forms. In the model discussed here, we have found the
simplest among the scenarios proposed in \cite{frme,cukusk}.
Further work is necessary to determine
the most
appropriate form of $\Sigma$ to interpret real glass experiments.
Our results
however, open the door for the  systematic study, starting from a mean-field
level, of aging effects in real glasses.

\vspace{1 cm}

It is a pleasure to thank C. de Dominicis, M. M\'ezard, G. Parisi for
interesting discussions and suggestions.

\begin{figure}
\vspace*{23cm}
\hbox to
\hsize{\hspace*{-4cm}\includegraphics{FHfig2.ps}\hspace*{4cm}}
\vspace*{-15cm}
\caption{}
\label{fig:exact}
\end{figure}

\section*{Figure caption}

Figure 1.
A sketch of the correlation and the response functions for
$g=\tilde{q}=1$, $T=.2<T_c$.
%Effects of aging in correlation and response functions.
(a). Normalized correlation function $C(t_w+\tau,t_w)/\tq$ and integrated
response function $m(t_w+\tau,t_w) = (T/\tq )\int_0^{t_w}r(t_w+\tau,s)\d s$
as functions of $\tau$, for different waiting times $t_w$.  If TTI and
the FDT held, these functions would be equal up to a constant shift and
independent of $t_w$.
(b). $m$ plotted against $C$ for different waiting times.  For large
values of $C$ (short times), the FDT holds, and the curves approach a
straight line of unit slope which intersects the $C$ axis at $q$.  The
deviation from this line at smaller $C$ (long times) indicates off-equilibrium
behaviour.
In the region $C<q$ the behaviour predicted by the asymptotic solution would be
a straight line
with slope $u$, starting from the origin.
The times displayed in the figure are very short compared to the ones needed
to see  the asymptotic behaviour.


\begin{thebibliography}{99}
\bibitem{AGESWE}
L.~Lundgren, P.~Svedlindh, P.~Nordblad and O.~Beckman,
Phys. Rev. Lett. {\bf 51} (1983) 911;
P.~Nordblad, P.~Svedlindh, L.~Lundgren, and L.~Sandlund,
Phys. Rev. {\bf B33} (1986) 645;

\bibitem{AGESG}
R.~V.~Chamberlin, G.~Mozurkevich and R.~Orbach,
Phys. Rev. Lett. {\bf 52} (1984) 867;
R.~Hoogerbets, Wei-Li Luo and R.~Orbach,
Phys. Rev. {\bf B34} (1986) 1719;
M.~Alba, J.~Hamman, M.~Ocio and Ph.~Refrigier,
J. Appl. Phys. {\bf 61} (1987) 3683;
F.~Lefloch, J.~Hamman, M.~Ocio and E.~Vincent,
Europhys. Lett. {\bf 18} (1992) 647;

\bi{cuku1} L. F. Cugliandolo and J. Kurchan; Phys. Rev. Lett. {\bf 71}, 1
(1993).

\bi{frme}S. Franz, M. M\'ezard, Europhys. Lett. {\bf 26} (1994) 209;
 { Physica } {\bf A209} (1994) 1

\bi{cukusk} L. F. Cugliandolo and J. Kurchan, preprint cond-mat/9311016, Roma I
977.

\bi{struik} L.C.E. Struik; {\it Physical aging in amorphous polymers and other
materials} (Elsevier, Houston 1978).

\bi{binder} W.Paul, J. Baschnagel, in {\it Monte Carlo and molecular dynamics
simulations in polymer science}, K. Binder editor (Oxford University Press,
NY, 1994), and references therein.

\bi{boumez} J.P. Bouchaud, M. M\'ezard {J. Phys. I France} {\bf 4} (1994) 1109

\bi{mapa1}  E. Marinari, G.Parisi, F.Ritort, preprint hep-th/9405148,
ROM2F/94/15; preprint cond-mat/9406074, Roma 1027

\bi{kuri} L. F. Cugliandolo,  J. Kurchan, G.Parisi, F.Ritort,
preprint Univ.Roma I, 1038/94, ROM2F/94/27, cond-mat/9407086

\bi{hor} A. Crisanti, H. Horner, H.J. Sommers, Z. Phys. B {\bf 92} 257 (1993)

\bi{leut} E. Leutheusser Phys. Rev. A {\bf 29} (1994) 2265

\bi{gotze}A review on the mode coupling theory
of glassy transition can be found in:
W. G\"otze, p. 403 in "Liquid, freezing and glass transition", Les Houches
(1989) J.P. Hansen, D. Levesque, J. Zinn-Justin editors, North Holland

\bi{dasmaz} S.P. Das, G.F. Mazenko Phys. Rev. A {\bf 34} (1986) 2265;
B. Kim, Phys. Rev. A {\bf 46} 1992 (1992); G.F. Mazenko, J. Yeo, preprint
cond-mat/9402023.

\bi{amro}D.J. Amit, Roginsky, {\it J.Phys. A} {\bf 12} (1979) 689.

\bi{criso} A. Crisanti, H.J. Sommers, Z. Phys. B {\bf 87} 341 (1992)

\bi{mouwe} C.Y. Mou, P.B. Weichmann, Phys. Rev. Lett. {\bf 72} 2041 (1994);
 J.P. Dorety M.A. Moore, J.M. Kim, A.J.  Bray,  Phys. Rev. Lett.
{bf 72} (1994) 2041.

\bi{fh2} S. Franz, J. Hertz, in preparation.

\bi{ZJ} see e.g. J. Zinn-Justin, {\it Quantum Field Theory and Critical
Phenomena},
Oxford University Press 1989; M.V. Feigelman, A.M. Tsvelik, Sov. Phys. JETP
{\bf 56} (1982) 823; J. Kurchan, J. Phys. France {\bf 2}, 1333 (1992)


\bi{gotzetto} W. G\"otze, Z Phys. B {\bf 56} 139 (1984); {\bf 60}, 195 (1985)





\end{thebibliography}
\end{document}